\documentclass[prd,superscriptaddress,amsfonts,amssymb,amsmath,showpacs,twocolumn,nofootinbib]{revtex4-2}
\usepackage{bm}
\usepackage{amsfonts}
\usepackage{latexsym}
\usepackage{graphicx}
\usepackage{amsmath}
\usepackage{palatino}
\usepackage{xcolor} 
\usepackage{mathpazo}
\usepackage{xcolor}
\usepackage{rotating}
\usepackage{adjustbox}
\usepackage{tensor}
\usepackage{textcomp}
\usepackage{float}
\usepackage{booktabs}
\usepackage{dcolumn}
\usepackage[titletoc]{appendix}
\usepackage{booktabs}
\usepackage{multirow}
\usepackage{hyperref}
\hypersetup{colorlinks,citecolor=blue}
\usepackage{amsmath}
\usepackage{xcolor}
\usepackage{orcidlink}
\usepackage{epsfig}
\usepackage{caption}
\usepackage{subcaption}
\usepackage{commath}

\hypersetup{colorlinks,citecolor=blue}
\hypersetup{colorlinks=true,linkcolor=magenta,filecolor=magenta,    urlcolor=blue}

\def\be{\begin{equation}}
\def\ee{\end{equation}}
\def\bea{\begin{eqnarray}}
\def\eea{\end{eqnarray}}

\begin{document}

\title{Constraints on the Modified Emergent Dark Energy Model Using DESI DR2 BAO}

\author{Himanshu Chaudhary}
\email{himanshu.chaudhary@ubbcluj.ro,\\
himanshuch1729@gmail.com}
\affiliation{Department of Physics, Babeș-Bolyai University, Kogălniceanu Street, Cluj-Napoca, 400084, Romania}
\affiliation{Research Center of Astrophysics and Cosmology, Khazar University, Baku, AZ1096, 41 Mehseti Street, Azerbaijan}

\author{Dhruba Jyoti Gogoi}
\email{moloydhruba@yahoo.in}
\affiliation{Department of Physics, Madhabdev University, Narayanpur, Lakhimpur 784164, Assam, India}

\author{Salvatore Capozziello}
\email{capozziello@na.infn.it}
\affiliation{Dipartimento di Fisica ``E. Pancini", Universit\`a di Napoli ``Federico II", Complesso Universitario di Monte Sant’ Angelo, Edificio G, Via Cinthia, I-80126, Napoli, Italy,}
\affiliation{Istituto Nazionale di Fisica Nucleare (INFN), sez. di Napoli, Via Cinthia 9, I-80126 Napoli, Italy,}
\affiliation{Scuola Superiore Meridionale, Via Mezzocannone 4, I-80134, Napoli, Italy.} 

\author{G. Mustafa}
\email{gmustafa3828@gmail.com}
\affiliation{Department of Physics,
Zhejiang Normal University, Jinhua 321004, People’s Republic of China}

\begin{abstract}
In this work, we investigated the modified emergent dark energy model using the baryon acoustic oscillation data from the Dark Energy Spectroscopic Instrument Data Release 2, together with CMB information from the {\tt LoLLiPoP} (low-$\ell$) and {\tt HiLLiPoP} (high-$\ell$) likelihoods based on the latest Planck NPIPE PR4 data release, along with the Planck PR4 lensing and ACT DR6 lensing likelihoods, as well as three different Type Ia supernova datasets (Pantheon$+$, DES-Dovekie, and Union3). Our analysis shows that the MEDE model remains largely consistent with the standard $\Lambda$CDM cosmology. Although the model allows small deviations from $\Lambda$CDM, the inclusion of SNe~Ia datasets drives the parameters toward values consistent with the standard cosmological model within the $1\sigma$ level. In particular, the parameter $\alpha$ does not show any statistically significant deviation from $\alpha=0$. We further find that the MEDE model does not significantly alleviate the current $H_0$ and $S_8$ tensions. The inferred values of $H_0$ remain lower than the SH0ES measurement, while the sound horizon scale $r_d$ stays nearly unchanged compared to the $\Lambda$CDM prediction. Similarly, although the MEDE model predicts slightly lower values of $S_8$, the reduction is not sufficient to ease the tension with weak-lensing observations. The evolution of the cosmological quantities closely follows that of $\Lambda$CDM, with deviations generally remaining below the $0.5\sigma$ level. Moreover, the CMB + DESI DR2 dataset combination favors a full phantom regime, whereas the inclusion of SNe~Ia datasets shifts the evolution toward a full quintessence regime. Finally, the $\Delta\chi^2_{\rm MAP}$ and $\Delta{\rm AIC}$ analyses indicate a slightly better fit of the MEDE model to the observational data, whereas the logarithmic Bayes factor shows inconclusive evidence in favour of the MEDE model over the $\Lambda$CDM model. Despite its overall consistency with the standard cosmological scenario, the MEDE model does not show the phantom-crossing behavior suggested by recent DESI DR2 observations.
\end{abstract}

\maketitle

\section{Introduction}\label{sec_1}
The $\Lambda$CDM model, commonly known as the concordance paradigm of modern cosmology, has achieved remarkable success in describing the statistical properties of the large-scale structure , the cosmic microwave background anisotropies, and the background expansion of the universe. It postulates a spatially flat universe dominated by cold dark matter (CDM) and a positive cosmological constant ($\Lambda$), with the latter being responsible for the observed late-time cosmic acceleration \cite{Planck:2018vyg,SupernovaSearchTeam:1998fmf, Copeland:2006wr,Sahni:1999gb}. Despite its unparalleled empirical robustness, the $\Lambda$CDM framework is not without profound theoretical challenges and growing observational tensions \cite{Park:2017xbl,Perivolaropoulos:2021jda,Bullock:2017xww,Hu:2023jqc}. 

From a theoretical standpoint, the model suffers from the cosmological constant problem arising from the striking discrepancy between the observed value of $\Lambda$ and the vastly larger vacuum energy density predicted by quantum field theory, posing a severe fine-tuning challenge \cite{Tian:2019enx,SolaPeracaula:2024nsz,Diaz-Pachon:2024nsq,Jain:2014wsa}. Equally puzzling is the coincidence problem, which questions why the energy densities of dark energy and dark matter are of approximately the same order of magnitude only in the current cosmological epoch \cite{Perivolaropoulos:2021jda,Velten:2014nra}. On the observational front, the standard model struggles to reconcile the persistent Hubble ($H_0$) tension and the structure growth ($\sigma_8$ or $S_8$) tension, which highlight statistically significant inconsistencies between early-universe predictions and late-universe local measurements \cite{Bhattacharyya:2018fwb,Pandey:2019plg,Hu:2023jqc,DiValentino:2021izs}. In response to these challenges, a plethora of theoretical frameworks have been proposed, including dynamical dark energy models \cite{Yang:2021eud,Sharma:2025qmv,Dinda:2024kjf,Chevallier:2000qy,Jassal:2004ej}, interacting dark energy scenarios \cite{Giare:2024smz,Li:2025owk,Clemson:2011an,Li:2014eha,Xia:2016vnp}, and modifications to general relativity \cite{Clifton:2011jh,Shankaranarayanan:2022wbx,Dolgov:2003px,KumarSharma:2022qdf,DeFelice:2010aj,Tsujikawa:2007xu,Sharma:2019yix,Sharma:2020vex,Sharma:2023vme}.

Recently, the landscape of precision cosmology has been catalyzed by the release of the Dark Energy Spectroscopic Instrument (DESI) Data Release 2 (DR2). Motivated by the DESI DR2 observations, which show a tantalizing preference for a dynamical dark energy equation of state over a static cosmological constant at a significance level ranging from 2.8$\sigma$ to 3.2$\sigma$, depending on the Type Ia supernova sample used, the community has been actively reinvestigating phenomenological and emergent dark energy frameworks. A compelling alternative to the standard $\Lambda$CDM model is the Generalized Emergent Dark Energy (GEDE)~\cite{Li:2020ybr} and Modified Emergent Dark Energy (MEDE)~\cite{benaoum2022modified} models. In such emergent scenarios, dark energy has no effective presence in the early universe, allowing standard radiation and matter-dominated eras to proceed unaltered, but naturally ``emerges'' at late times to drive cosmic acceleration. These models are phenomenologically rich, observationally testable, and theoretically motivated by the desire to address both the coincidence problem and the contemporary $H_0$ and $S_8$ tensions.

In this paper, we go a step further in our investigation of the MEDE framework by performing a robust, high-precision Markov Chain Monte Carlo analysis utilizing the most up-to-date and complementary cosmological datasets. In particular, we combine the Baryon Acoustic Oscillation measurements from DESI Data Release~2 with Cosmic Microwave Background observations from {\it Planck PR4} and ACT. To rigorously test the late-time expansion history and break parameter degeneracies, we systematically combine these probes with three independent Type Ia Supernovae catalogs: the Pantheon$+$ compilation, the DES-Dovekie sample, and the Union3 compilation.

Our primary objective is to evaluate whether the MEDE model can successfully alleviate the $H_0$ and $S_8$ tensions while remaining consistent with the combined observational data. Furthermore, we aim to definitively test whether the MEDE framework shows the specific dynamical dark energy behaviors such as the phantom-crossing (Quintom-B) scenario recently favored by the DESI DR2 background expansion measurements. By tracking the deviation parameter ($\alpha$), we assess the statistical viability of the MEDE model against the standard $\Lambda$CDM paradigm using Bayesian evidence.

The paper is organized as follows: In Section \ref{sec_2}, we provide an overview of the theoretical background and the mathematical formulation of the Modified Emergent Dark Energy model. This Section also details the recent observational datasets and the MCMC methodology utilized in this work. In Section \ref{sec_4}, we present the constraints obtained on the cosmological parameters, discuss the implications for the $H_0$ and $S_8$ tensions, and analyze the dynamical behavior of the dark energy equation of state $w(z)$. Finally, we summarize our conclusions and findings.

\section{Theoretical background}\label{sec_2}
The theoretical framework of the MEDE scenario is constructed within the standard homogeneous and isotropic cosmological background described by the spatially flat Friedmann-Lemaître-Robertson-Walker (FLRW) metric,
\begin{equation}
ds^2=-dt^2+a^2(t)(dx^2+dy^2+dz^2),
\end{equation}
where $a(t)$ is the cosmic scale factor. Assuming General Relativity as the underlying gravitational theory and considering a non--interacting cosmic fluid sector, the cosmic expansion history is governed by the Friedmann equation,
\begin{equation}
H^2(z)=H_0^2\left[\Omega_{r0}(1+z)^4+\Omega_{m0}(1+z)^3+\tilde{\Omega}_{\rm DE}(z)\right],
\end{equation}
where $\Omega_{r0}$ and $\Omega_{m0}$ denote the present radiation and matter density parameters, respectively, while $\tilde{\Omega}_{\rm DE}(z)\equiv \rho_{\rm DE}(z)/\rho_{\rm crit,0}$ represents the normalized dark energy density.

Instead of directly parametrizing the dark energy equation of state, the MEDE framework parametrizes the dark energy density evolution through
\begin{equation}
\tilde{\Omega}_{\rm DE}(z)=\Omega_{{\rm DE},0}\,G(z),
\end{equation}
where $G(z)$ is a generic function satisfying $G(0)=1$. Following the emergent dark energy approach, the generalized function is chosen as
\begin{equation}
G(z)=1-\tanh\!\left[\log_{10}g(z)\right],
\end{equation}
with $g(z)$ being a continuous function satisfying $g(0)=1$. This construction naturally generalizes the Phenomenologically Emergent Dark Energy (PEDE) model \cite{li2019simple}, which is recovered for $g(z)=1+z$, while the $\Lambda$CDM cosmology corresponds to the trivial choice $g(z)=1$. Hence, the MEDE framework provides a unified phenomenological platform capable of describing several dark energy scenarios through different choices of $g(z)$.

The dark energy equation of state can be obtained from the conservation equation,
\begin{equation}
\dot{\rho}_{\rm DE}+3H(1+w_{\rm DE})\rho_{\rm DE}=0,
\end{equation}
leading to
\begin{equation}
w_{\rm DE}(z)
=-1+\frac{1}{3}(1+z)\frac{d\ln \tilde{\Omega}_{\rm DE}}{dz}.
\end{equation}
To determine the functional form of $g(z)$, one imposes the condition
\begin{equation}
\frac{1+z}{g(z)}\frac{dg(z)}{dz}=\alpha,
\end{equation}
where $\alpha$ is a constant parameter controlling the deviation from the standard cosmological constant scenario. Solving this equation gives
\begin{equation}
g(z)=(1+z)^{\alpha}.
\end{equation}
Consequently, the normalized dark energy density becomes
\begin{equation}
G(z)=1-\tanh\!\left[\alpha\log_{10}(1+z)\right],
\end{equation}
and the dimensionless Hubble expansion rate takes the form
\begin{equation}
\begin{aligned}
E(z) &= \frac{H(z)}{H_0} = \Bigg[
\Omega_{bc}(1+z)^3 + \Omega_{\gamma}(1+z)^4 + \Omega_{k}(1+z)^2  \\
&\quad + \Omega_{{\rm DE}}
\left(
1-\tanh\!\left[\alpha \log_{10}(1+z)\right]
\right)
\Bigg]^{1/2}.
\end{aligned}
\end{equation}

Here, $\Omega_{bc} = \Omega_b + \Omega_c$, and $\Omega_{\gamma}$, $\Omega_{K}$, and $\Omega_{\rm DE}$ refer to the energy densities of radiation, curvature, and dark energy, respectively. In our analysis, we consider the spatially flat case, i.e., $\Omega_{K}=0$. The corresponding dark energy equation of state evolves as follows.

\begin{equation}
w_{\rm DE}(z)
=-1-\frac{\alpha}{3\ln(10)}
\left[1+\tanh\!\left(\alpha\log_{10}(1+z)\right)\right].
\end{equation}

The parameter $\alpha$ plays a crucial role in determining the dynamical nature of dark energy. Specifically, $\alpha=0$ reproduces the standard $\Lambda$CDM cosmology, while $\alpha=1$ recovers the original PEDE model. Furthermore, positive values of $\alpha$ correspond to phantom-like dark energy $(w_{\rm DE}<-1)$, whereas negative values lead to quintessence-like behavior $(w_{\rm DE}>-1)$. Interestingly, irrespective of the value of $\alpha$, the model asymptotically approaches $w_{\rm DE}\rightarrow -1$ in the far future $(z\rightarrow -1)$, mimicking a de Sitter phase.

\begin{figure*}
\begin{subfigure}{.49\textwidth}
\includegraphics[width=\linewidth]{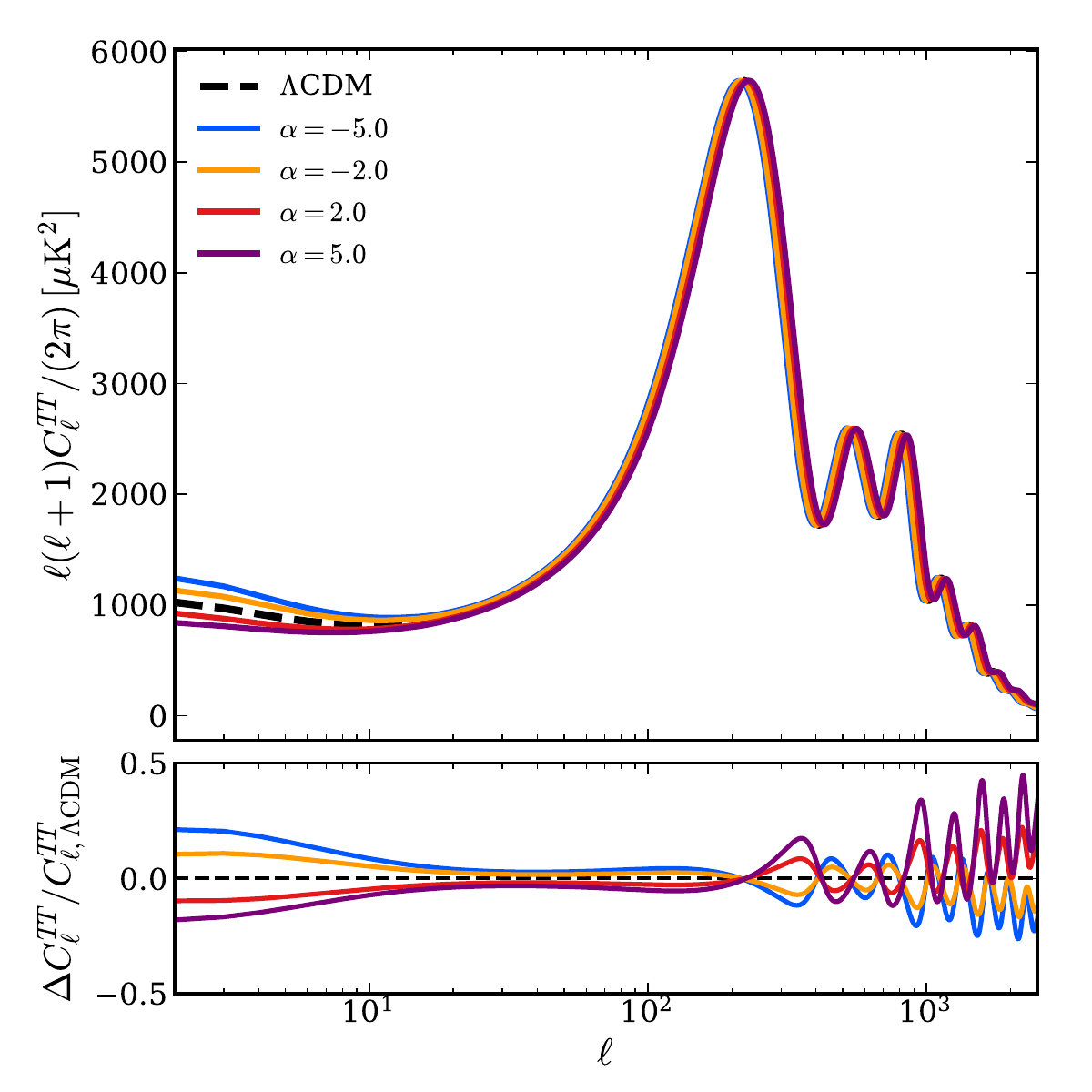}
\end{subfigure}
\hfil
\begin{subfigure}{.49\textwidth}
\includegraphics[width=\linewidth]{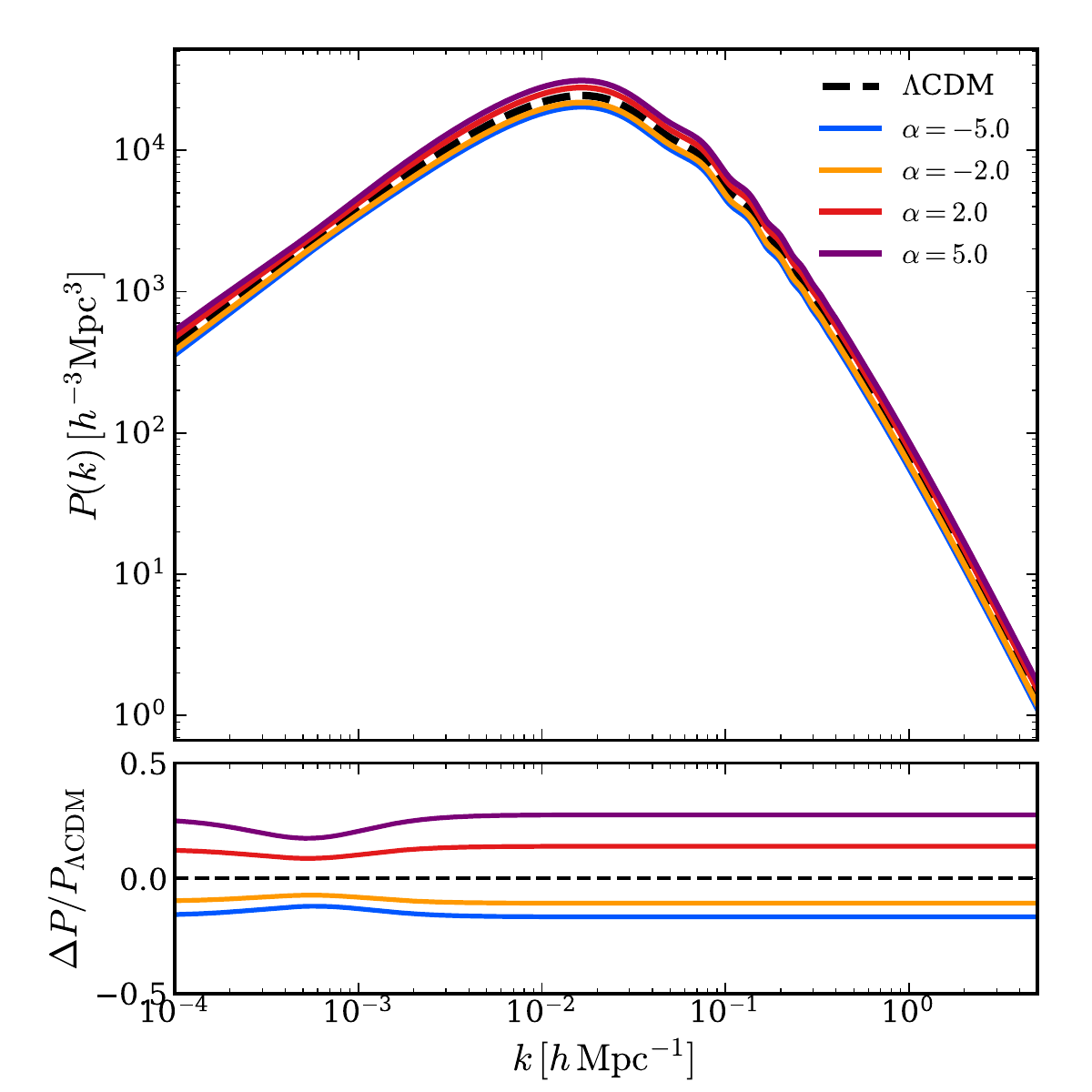}
\end{subfigure}
\caption{This figure shows the CMB temperature power spectrum, $C_\ell^{TT}$ (left panel), together with the corresponding fractional difference, $\Delta C_\ell^{TT}/C_{\ell,\Lambda{\rm CDM}}^{TT}$ (lower-left panel). The right panel shows the linear matter power spectrum, $P(k)$, and its corresponding fractional difference, $\Delta P/P_{\Lambda{\rm CDM}}$ (lower-right panel). The black dashed curve shows the $\Lambda$CDM model, whereas the colored curves correspond to different values of the model parameter $\alpha$, with all other cosmological parameters fixed.}\label{fig_TT_Matter}
\end{figure*}

Before moving to the main MCMC analysis, we first show the impact of different values of $\alpha$ on the CMB temperature anisotropy and the linear matter power spectrum in Fig.~\ref{fig_TT_Matter}. The upper-left panel shows that varying $\alpha$ produces only modest changes in the CMB temperature power spectrum. The largest deviations occur at low multipoles ($\ell\lesssim20$), where cosmic variance is also largest, while the positions of the acoustic peaks remain nearly unchanged. Small variations in the peak amplitudes are visible around the first acoustic peak, indicating that the model primarily affects the late-time evolution of the Universe rather than the physics of recombination.

The lower-left panel shows the fractional difference relative to the $\Lambda$CDM model. The deviations remain below the $\sim50\%$ level over the full multipole range and are most pronounced at low multipoles and at high multipoles ($\ell\gtrsim800$), whereas the intermediate multipole region remains very close to the standard cosmological prediction. Positive values of $\alpha$ generally enhance the high-$\ell$ oscillatory residuals, while negative values suppress them.

The upper-right panel shows the corresponding linear matter power spectrum. Unlike the CMB spectrum, the matter power spectrum shows a systematic amplitude shift over a broad range of scales. Increasing $\alpha$ enhances the amplitude of $P(k)$, whereas negative values reduce it with respect to the $\Lambda$CDM prediction. The overall shape of the spectrum and the turnover scale remain almost unchanged, indicating that the model primarily modifies the growth amplitude rather than the characteristic scale of matter clustering.

This behaviour is more clearly seen in the lower-right panel, which shows the fractional difference with respect to the $\Lambda$CDM model. The deviations are nearly scale independent on large scales and become slightly larger around the turnover region before remaining almost constant on smaller linear scales. This indicates that the parameter $\alpha$ mainly affects the growth of cosmic structures by changing the overall amplitude of matter clustering, while leaving the overall shape of the matter power spectrum largely unchanged.

\begin{figure}
\centering
\includegraphics[scale=0.43]{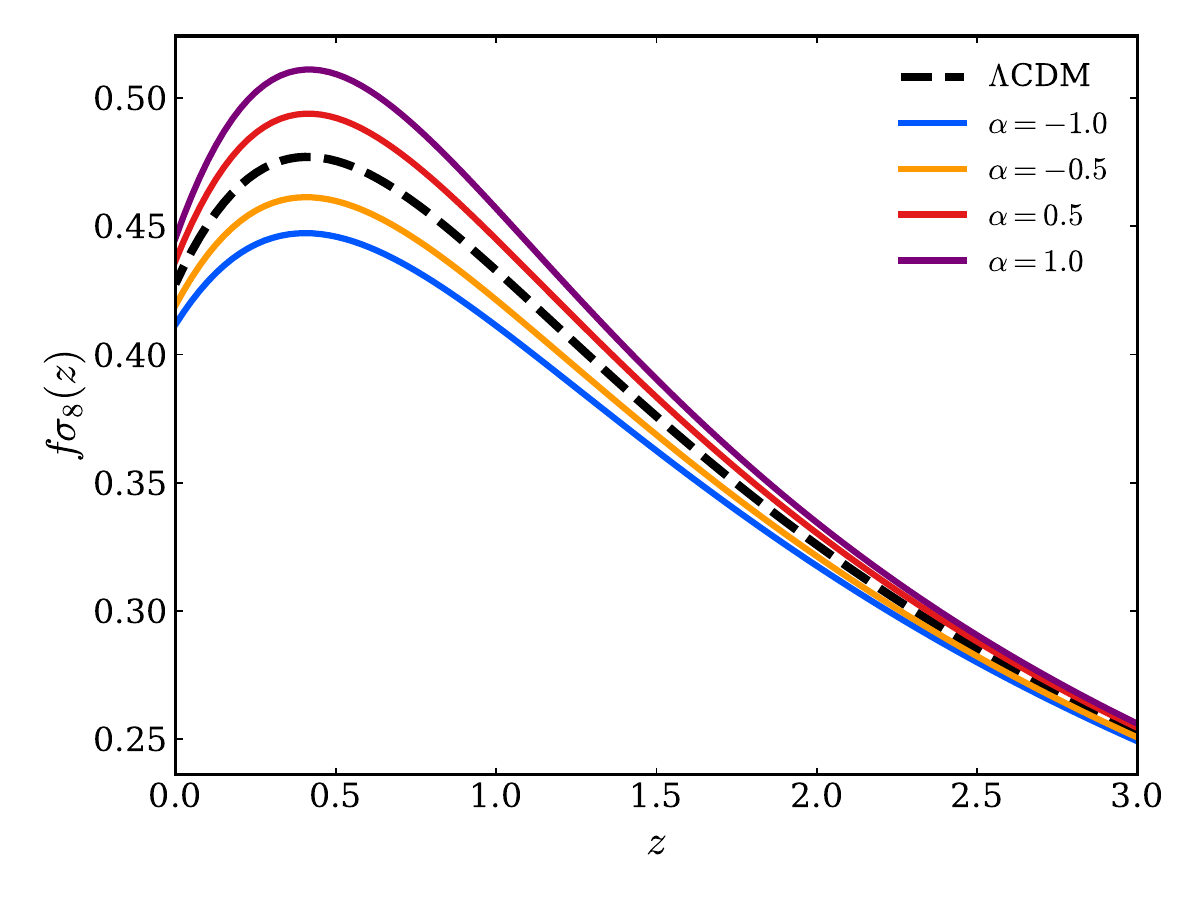}
\caption{This figure shows the evolution of the growth-rate observable $f\sigma_8(z)$ as a function of redshift for different values of the model parameter $\alpha$. The black dashed curve corresponds to the $\Lambda$CDM model, while the coloured curves show the predictions for different values of $\alpha$.}\label{fig_fsigma8}
\end{figure}

To further examine the impact of the parameter $\alpha$ on the growth of cosmic structures, we also show the evolution of the growth-rate $f\sigma_8(z)$ in Fig.~\ref{fig_fsigma8}. We find that changing $\alpha$ mainly modifies the amplitude of $f\sigma_8(z)$ while preserving its overall redshift evolution. Positive values of $\alpha$ predict a larger growth rate than the $\Lambda$CDM model, whereas negative values suppress the growth of structures. The differences are most pronounced at low redshifts, where dark energy becomes dynamically important, and gradually decrease toward higher redshifts, where all models approach the standard cosmological evolution. This behaviour is fully consistent with the trends observed in the matter power spectrum, confirming that the parameter $\alpha$ primarily affects the late-time growth of cosmic structures rather than the underlying shape of the matter distribution.

\section{Dataset and Methodology}\label{sec_3}
To constrain the cosmological parameters of the MEDE model, we perform a Bayesian analysis using the publicly available package {\tt Cobaya}\footnote{\url{https://github.com/CobayaSampler/cobaya}} \cite{torrado2021cobaya}, interfaced with the Boltzmann solver {\tt CAMB}\footnote{\url{https://github.com/cmbant/CAMB.git}} \cite{lewis2000efficient,howlett2012cmb}. The parameter space of the MEDE model is explored through Markov Chain Monte Carlo (MCMC) sampling \cite{lewis2002cosmological,lewis2013efficient,neal2005taking}, while the convergence of the chains is assessed using the Gelman-Rubin criterion \cite{gelman1992inference}, requiring $R-1<0.01$. The resulting chains are analyzed using the {\tt GetDist\footnote{\url{https://github.com/cmbant/getdist.git}}} package \cite{lewis2025getdist}.

For the baseline $\Lambda$CDM model, the set of free cosmological parameters is given by $\bm{\theta}_{\Lambda{\rm CDM}}
=
\{
\Omega_{\rm cdm},
\Omega_{\rm b},
100\,\theta_{\rm MC},
\ln(10^{10}A_{\rm s}),
n_{\rm s},
\tau
\}.$
For the MEDE model, the parameter space is extended as $\bm{\theta}_{\rm MEDE}
=
\{
\bm{\theta}_{\Lambda{\rm CDM}},
\alpha
\},$
where $\alpha$ characterizes the evolution of the emergent dark energy sector.

In addition to parameter estimation, we perform Bayesian model comparison between the MEDE and $\Lambda$CDM cosmologies. The Bayesian evidence, $\mathcal{Z}$, is computed using {\tt \texttt{MCEvidence}\footnote{\url{https://github.com/yabebalFantaye/MCEvidence.git}}} \cite{heavens2017marginal,heavens2017no}, through the Cobaya interface available in the {\tt wgcosmo} repository\footnote{\url{https://github.com/williamgiare/wgcosmo.git}}. The relative preference between the two models is quantified through the logarithmic Bayes factor, $\ln B
=
\ln \mathcal{Z}_{\Lambda{\rm CDM}}
-
\ln \mathcal{Z}_{\rm MEDE}.$ The statistical interpretation of $\ln B$ follows the revised Jeffreys scale \cite{kass1995bayes,trotta2008bayes}, where $\ln B<1$ indicates inconclusive evidence, $1\leq \ln B<2.5$ weak evidence, $2.5\leq \ln B<5$ moderate evidence, $5\leq \ln B<10$ strong evidence, and $\ln B\geq10$ decisive evidence in favor of the $\Lambda$CDM model. While a negative value of $\ln B$ shows that the MEDE model is favored over the $\Lambda$CDM model.

We also calculate the difference in the minimum chi-square, defined as $\Delta\chi^2_{\rm MAP}=\chi^2_{\Lambda{\rm CDM}}-\chi^2_{\rm MEDE}$, where positive values indicate that the MEDE model provides a better fit to the data than the $\Lambda$CDM model. Furthermore, we compute the difference in the Akaike Information Criterion (AIC) \cite{liddle2007information,tan2012reliability}, defined as $\Delta{\rm AIC}={\rm AIC}_{\Lambda{\rm CDM}}-{\rm AIC}_{\rm MEDE}$, with ${\rm AIC}=\chi^2_{\rm MAP}+2k$, where $k$ denotes the number of free parameters of the corresponding model. According to this convention, positive values of $\Delta{\rm AIC}$ indicate a preference for the MEDE model after accounting for its additional free parameter, whereas negative values favor the $\Lambda$CDM model.

To constrain the parameters of the MEDE model, we consider several combinations of late and early Universe observations, including Baryon Acoustic Oscillation, Type~Ia Supernova, and Cosmic Microwave Background datasets, which are described below.

\begin{itemize}
     \item \textbf{Baryon Acoustic Oscillation :} First, we consider the Baryon Acoustic Oscillation (BAO) measurements from the Dark Energy Spectroscopic Instrument (DESI) Data Release~2\footnote{\url{https://github.com/CobayaSampler/bao_data.git}} (DR2) \cite{abdul2025desi}. The dataset includes measurements from multiple tracers, namely the Bright Galaxy Sample (BGS), Luminous Red Galaxies (LRG), Emission Line Galaxies (ELG), Quasars (QSO), and the Lyman-$\alpha$ forest, spanning the redshift range $0.3 \leq z \leq 2.33$. The DESI DR2 BAO likelihood is implemented using the compressed distance measurements $D_M/r_d$, $D_H/r_d$, and $D_V/r_d$, where $D_H(z)=c/H(z)$ is the Hubble distance, $D_M(z)=c\int_0^z dz'/H(z')$ is the transverse comoving distance, $D_V(z)=[zD_M^2(z)D_H(z)]^{1/3}$ is the isotropic volume-averaged distance, and $r_d$ denotes the sound horizon at the baryon drag epoch.
     
     \item \textbf{Type Ia Supernova :}
     Second, we consider three different Type~Ia supernova (SNe~Ia) compilations. The first is the Pantheon$+$\footnote{\url{https://github.com/PantheonPlusSH0ES/DataRelease.git}} sample \cite{scolnic2022pantheon}, containing 1,701 light curves from 1,550 SNe~Ia in the redshift range $0.001 \leq z \leq 2.26$. Following the standard Pantheon$+$ analysis \cite{brout2022pantheon}, we exclude supernovae with $z<0.01$ to reduce the impact of peculiar velocity systematics. We also use the DES-Dovekie\footnote{\url{https://github.com/des-science/DES-SN5YR.git}} compilation \cite{popovic2025dark}, which includes 1,820 photometrically calibrated SNe~Ia covering $0.02 \leq z \leq 1.14$. Finally, we consider the Union3\footnote{\url{https://github.com/rubind/union3_release}} sample \cite{rubin2025union}, consisting of 2,087 SNe~Ia spanning the redshift interval $0.05 < z < 2.26$. These SNe~Ia datasets provide complementary constraints on the late-time expansion history of the Universe.

     \item \textbf{Cosmic Microwave Background:} Finally, we consider Cosmic Microwave Background (CMB) observations from both {\it Planck} and ACT. For the {\it Planck} data, we use the low-$\ell$ temperature and polarization likelihoods ({\tt TT} and {\tt EE}), together with the NPIPE-based low-$\ell$ {\tt LoLLiPoP\footnote{\url{https://github.com/planck-npipe/lollipop.git}}} and high-$\ell$ {\tt HiLLiPoP\footnote{\url{https://github.com/planck-npipe/hillipop.git}}} likelihoods \cite{tristram2024cosmological}. We further include the {\it Planck} PR4 lensing likelihood\footnote{\url{https://github.com/carronj/planck_PR4_lensing.git}} \cite{carron2022cmb,carron2022planck} and the ACT DR6 lensing likelihood\footnote{\url{https://github.com/ACTCollaboration/act_dr6_lenslike.git}} \cite{madhavacheril2024atacama,qu2024atacama}.
\end{itemize}
The priors chosen for the $\Lambda$CDM and MEDE models are summarized in Table~\ref{tab_1}.


{
\renewcommand{\arraystretch}{1}
\begin{table}[t] 
    \centering
    \begin{tabular}{|lll|}
    \hline
    parametrization & parameter & prior\\  
    \hline 
    $\mathbf{\Lambda}$\textbf{CDM} & $\Omega_\mathrm{cdm}h^2$ & $\mathcal{U}[0.001, 0.99]$ \\   
    & $\Omega_\mathrm{b}h^{2}$ & $\mathcal{U}[0.005, 0.1]$ \\
    & $100\theta_\mathrm{MC}$ & $\mathcal{U}[0.5, 10]$ \\
    & $\ln(10^{10} A_\mathrm{s})$ & $\mathcal{U}[1.61, 3.91]$ \\
    & $n_\mathrm{s}$ & $\mathcal{U}[0.8, 1.2]$ \\
    & $\tau$ & $\mathcal{U}[0.01, 0.8]$ \\
    \hline 
    \textbf{Emergent} & $\alpha$ & $\mathcal{U}[-10, 10]$ \\
    \hline
    \end{tabular}
    \caption{ Parameters and priors used in the analysis. Here $\mathcal{U}[{\rm min, max}]$ denotes a uniform prior over the specified range.}\label{tab_1}
\end{table}
}


\begin{figure*}
\centering
\includegraphics[scale=0.42]{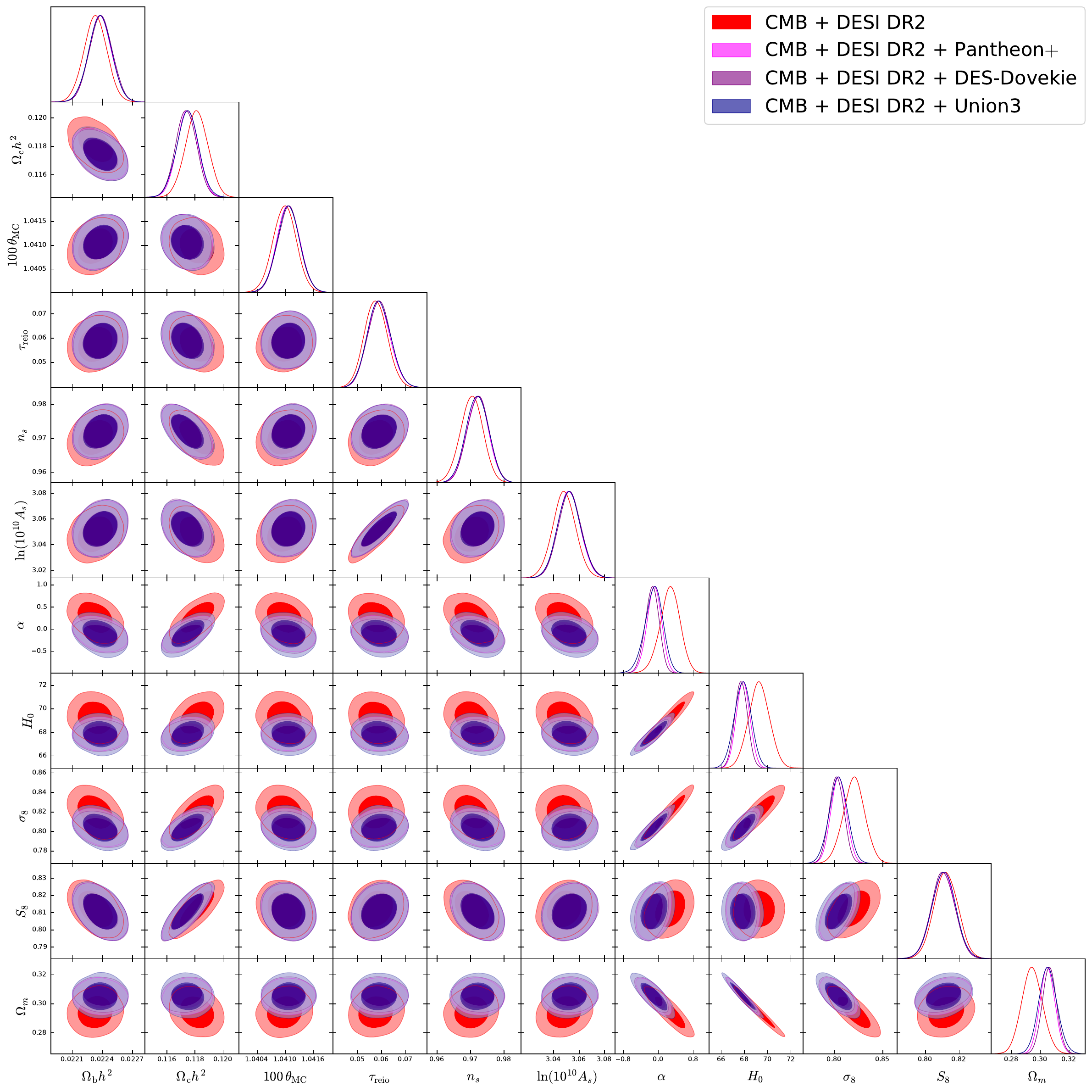}
\caption{The figure shows the corner plot of the MEDE model at 68\% ($1\sigma$) and 95\% ($2\sigma$) confidence levels using DESI DR2 combined with CMB and SNe~Ia datasets (Pantheon$+$, DES-Dovekie, and Union3), shown as superimposed contours for the different dataset combinations.}\label{fig_1}
\end{figure*}

\begin{figure*}
\centering
\includegraphics[scale=0.41]{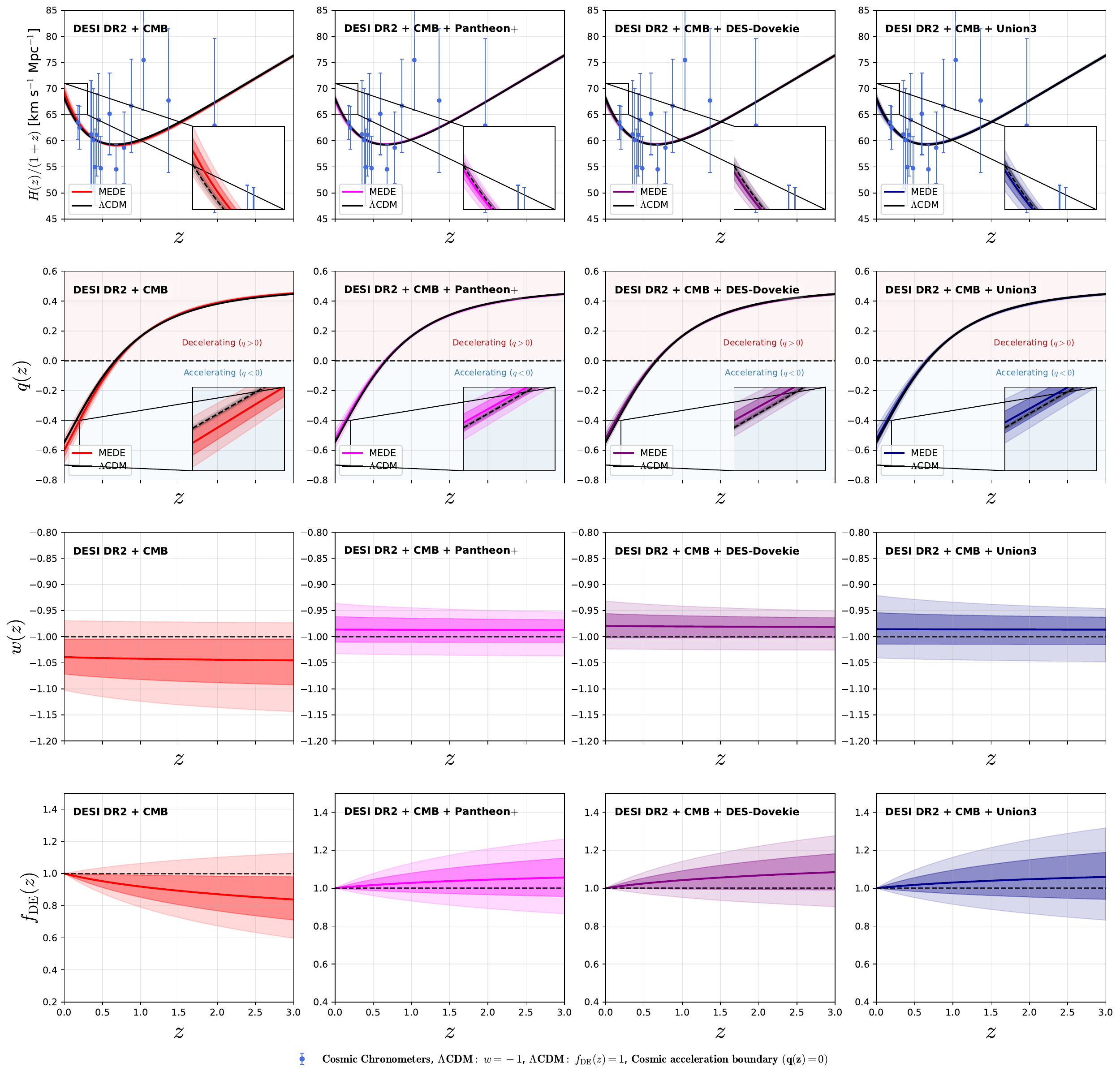}
\caption{The figure shows the evolution of $H(z)/(1+z)$ (first row), $q(z)$ (second row), $w(z)$ (third row), and $f_{\mathrm{DE}}(z)$ (fourth row) for the MEDE and $\Lambda$CDM models as functions of $z$. The solid lines represent the mean predictions, while the light and dark shaded regions correspond to the $1\sigma$ and $2\sigma$ confidence intervals, respectively.}\label{fig_3}
\end{figure*}

\begin{figure}
\centering
\includegraphics[scale=0.54]{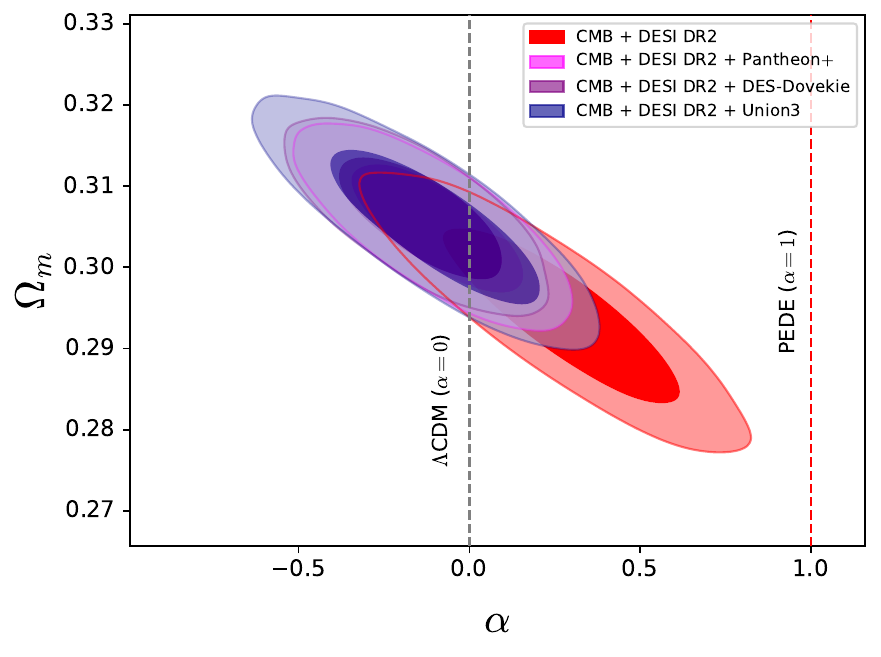}
\caption{The figure shows the 2D marginalized confidence contours of the $\{\alpha-\Omega_m\}$ plane of the MEDE model at 68\% ($1\sigma$) and 95\% ($2\sigma$) confidence levels using DESI DR2 combined with CMB and SNe~Ia datasets (Pantheon$+$, DES-Dovekie, and Union3), shown as superimposed contours for the different dataset combinations.}\label{fig_2}
\end{figure}

\begin{figure}
\centering
\includegraphics[scale=0.36]{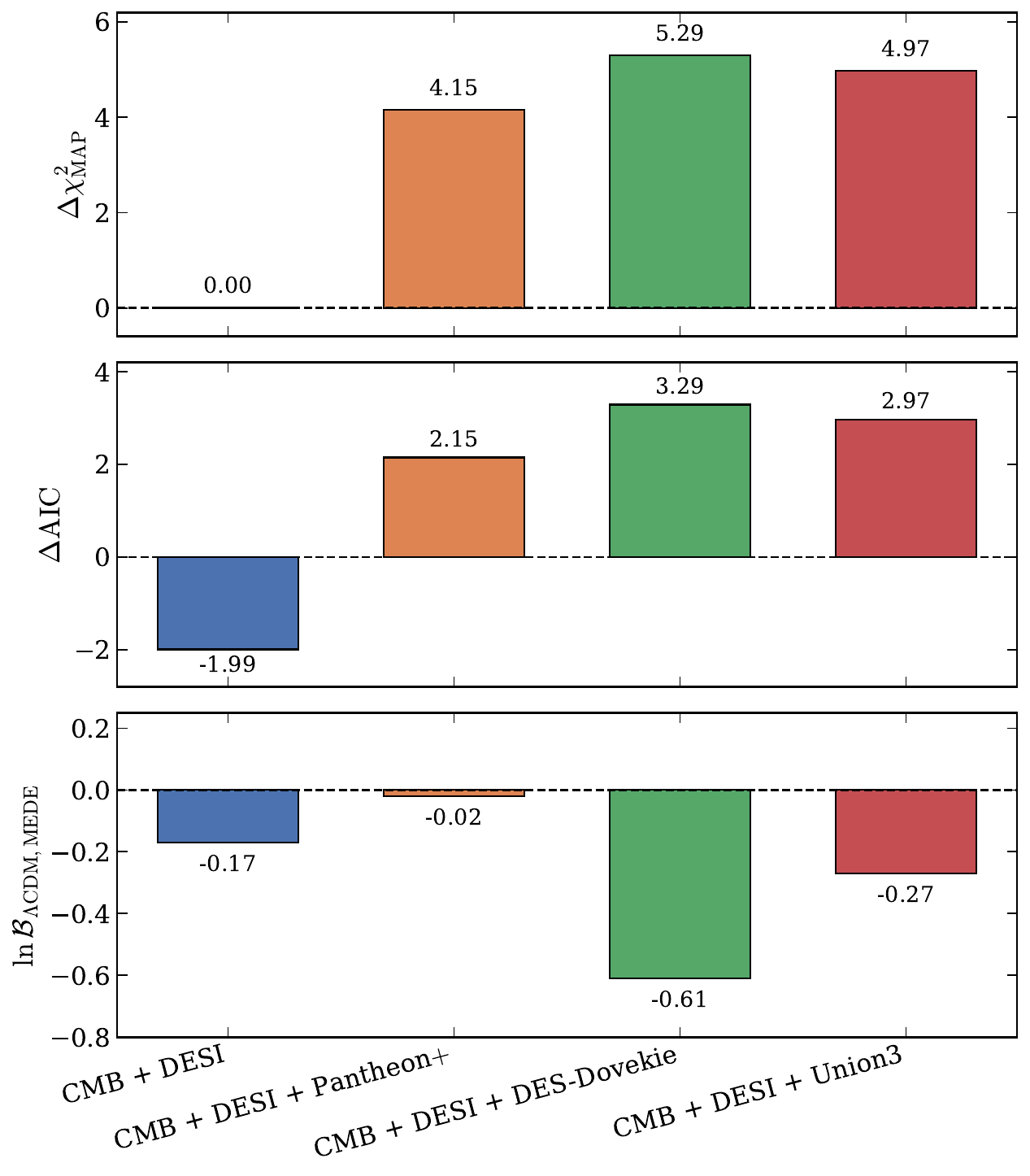}
\caption{The figure shows the Statistical comparison between the MEDE and $\Lambda$CDM models for the different dataset combinations. From top to bottom, the panels show $\Delta\chi^2_{\rm MAP}$, $\Delta{\rm AIC}$, and the logarithmic Bayes factor, $\ln\mathcal{B}_{{\rm MEDE},\Lambda{\rm CDM}}$.}\label{fig_stat}
\end{figure}
\begin{table*}[htbp]
\centering
\resizebox{\textwidth}{!}{%
\begin{tabular}{l@{\hspace{10pt}}c@{\hspace{10pt}}c@{\hspace{10pt}}c@{\hspace{10pt}}c}
\hline
\textbf{Parameters} 
& \textbf{CMB + DESI DR2} 
& \textbf{CMB + DESI DR2 + Pantheon$+$} 
& \textbf{CMB + DESI DR2 + DES-Dovekie} 
& \textbf{CMB + DESI DR2 + Union3} \\
\hline

$\Omega_{\mathrm{\text{b}}}h^2$ 
& $0.02233 \pm 0.00011$ 
& $0.02237 \pm 0.00011$
& $0.02238 \pm 0.00011$
& $0.02237 \pm 0.00011$ \\

$\Omega_{\mathrm{\text{cdm}}}h^2$ 
& $0.11812 \pm 0.00080$ 
& $0.11746 \pm 0.00074$
& $0.11737 \pm 0.00074$
& $0.11747 \pm 0.00077$ \\

$100\,\theta_{\mathrm{MC}}$ 
& $1.04098 \pm 0.00025$ 
& $1.04106 \pm 0.00024$
& $1.04107 \pm 0.00024$
& $1.04107 \pm 0.00024$ \\

$\tau_{\mathrm{reio}}$ 
& $0.0576 \pm 0.0048$ 
& $0.0589 \pm 0.0048$
& $0.0590 \pm 0.0049$
& $0.0589 \pm 0.0049$ \\

$n_{\mathrm{s}}$ 
& $0.9703 \pm 0.0034$ 
& $0.9721 \pm 0.0033$
& $0.9723 \pm 0.0033$
& $0.9720 \pm 0.0033$ \\

$\ln\bigl(10^{10}\,A_{\mathrm{s}}\bigr)$ 
& $3.0483 \pm 0.0090$ 
& $3.0522 \pm 0.0089$
& $3.0527 \pm 0.0090$
& $3.0522 \pm 0.0091$ \\

$\alpha$ 
& $0.27_{-0.21}^{+0.24}$ 
& $-0.097 \pm 0.17$
& $-0.14 \pm 0.16$
& $-0.11_{-0.19}^{+0.22}$ \\

\hline

$H_0 (\mathrm{km\,s^{-1}\,Mpc^{-1}})$ 
& $69.26 \pm 0.90$ 
& $67.88 \pm 0.59$
& $67.72 \pm 0.55$
& $67.85 \pm 0.73$ \\

$\sigma_{8}$ 
& $0.820 \pm 0.011$ 
& $0.8042 \pm 0.0080$
& $0.8022 \pm 0.0077$
& $0.8039 \pm 0.0094$ \\

$S_{8}$ 
& $0.8121 \pm 0.0069$ 
& $0.8107 \pm 0.0068$
& $0.8104 \pm 0.0069$
& $0.8109 \pm 0.0070$ \\

$\Omega_{\mathrm{m}}$ 
& $0.2943 \pm 0.0072$ 
& $0.3049 \pm 0.0052$
& $0.3062 \pm 0.0050$
& $0.3053 \pm 0.0063$ \\

$r_d\text{(Mpc)}$ 
& $147.65 \pm 0.20$ 
& $147.77 \pm 0.19$
& $147.79 \pm 0.19$
& $147.77 \pm 0.19$ \\

\hline

$\Delta{\chi^{2}_{\text{MAP}}}$
& 0.002
& 4.15
& 5.29
& 4.97 \\

$\Delta \rm AIC$
& -1.99
& 2.15
& 3.29
& 2.97 \\

$\ln B_{i,j}$ 
& $-0.17$
& $-0.02$
& $-0.61$
& $-0.27$\\

\hline
\end{tabular}
}
\caption{This table shows the constraints on the MEDE parameters, presenting the mean values and their uncertainties at the 68\% ($1\sigma$) confidence level, obtained using DESI DR2 combined with CMB and SNe~Ia datasets (Pantheon$+$, DES-Dovekie, and Union3).}\label{tab_2}
\end{table*}


\section{Results and Conclusions}\label{sec_4}
Fig.~\ref{fig_1} presents the marginalized constraints on the parameters of the MEDE model derived from the joint analysis of DESI DR2, CMB, and Type Ia supernova datasets, including Pantheon$+$, DES-Dovekie, and Union3. The two-dimensional (2D) contour panels show the correlations between different parameter pairs of the MEDE model, with the shaded inner and outer regions representing the 68\% ($1\sigma$) and 95\% ($2\sigma$) confidence intervals, respectively. The diagonal panels represent the corresponding 1D marginalized posterior distributions for the cosmological parameters of the model. The numerical constraints, reported as mean values together with their associated 68\% ($1\sigma$) uncertainties, are summarized in Table~\ref{tab_2} for the each combinations of DESI DR2, CMB, and SNe~Ia observations.

First, we discuss whether the MEDE model is capable of alleviating the $H_0$ tension after the inclusion of DESI DR2 data. In order to successfully address the $H_0$ tension, the inferred value of $H_0$ should be consistent with the SH0ES measurement obtained from the Cepheid-calibrated Type Ia supernova distance ladder, namely $(73.04 \pm 1.04),\mathrm{km,s^{-1},Mpc^{-1}}$~\cite{riess2022comprehensive}. At the same time, this should be achieved while reducing the sound horizon scale $r_d$ by approximately $7\%$~\cite{knox2020hubble,jedamzik2021reducing,vagnozzi2023seven}. It is important to note that such a reduction in $r_d$ generally requires a larger physical matter density, which can further worsen the $S_8 \equiv \sigma_8 \sqrt{\Omega_m/0.3}$ tension. As a consequence, attempts to alleviate the $H_0$ tension in this way may increase the disagreement between the predictions of the standard $\Lambda$CDM model and weak lensing (WL) observations~\cite{vagnozzi2023seven,jedamzik2021reducing}.

In the case of the MEDE model, we do not find any significant alleviation of the $H_0$ tension across the different dataset combinations. As shown in Table~\ref{tab_2}, the inferred values of $H_0$ remain within the range $H_0 \sim 67.72$–$69.26\ \mathrm{km,s^{-1},Mpc^{-1}}$, depending on the choice of datasets. Yet, it is important to note that, in the case of CMB + DESI DR2, the MEDE model predicts a higher value of $H_0$ than the value reported by the Planck Collaboration ($H_0 = 67.4 \pm 0.5~\mathrm{km\,s^{-1}\,Mpc^{-1}}$)~\cite{Planck:2018vyg}. The main reason for this can be understood from the evolution of the dark energy equation of state, $w(z)$. For the CMB + DESI DR2 combination, the MEDE model predicts a full phantom evolution (see the first column and third row of Fig.~\ref{fig_3}). As discussed in Refs.~\cite{Vagnozzi:2018jhn,Vagnozzi:2019ezj,Alestas:2020mvb,Banerjee:2020xcn,Lee:2022cyh,Colgain:2025nzf}, changing the dark energy equation of state from $w(z)=-1$ ($\Lambda$CDM) to $w(z)>-1$ (quintessence) generally prefers a lower value of $H_0$, whereas $w(z)<-1$ (phantom) prefers a higher inferred value of $H_0$. Accordingly, the CMB + DESI DR2 combination predicts a slightly higher value of $H_0$, whereas the remaining dataset combinations favor lower values because they shows a fully quintessence evolution. However, these values still show a noticeable deviation from the SH0ES measurement.

Indeed, one may misinterpret the fact that the model predicts a higher value of $H_0$ and conclude that it alleviates the Hubble tension. However, one should be careful with this interpretation. As is well known, the MEDE model belongs to the class of dynamical dark energy models, which modify the expansion history only at post-recombination times. Therefore, although the model predicts different values of $H_0$ for different dataset combinations, the sound horizon at the drag epoch, $r_d$, remains essentially unchanged. This can be clearly seen from the inferred values of $r_d$ predicted by the MEDE model, which remain in the range $147.65$--$147.79~\mathrm{Mpc}$ for all dataset combinations. Consequently, the model can predict either a higher or a lower value of $H_0$, yet the value of $r_d$ remains unchanged, and the model becomes inconsistent with BAO measurements~\cite{Bernal:2016gxb,Aylor:2018drw,Knox:2019rjx,Efstathiou:2021ocp,Jiang:2024xnu}.

For the MEDE model, we obtain $\sigma_8 \sim 0.804$–$0.820$ and $S_8 \sim 0.810$–$0.812$, depending on the choice of the datasets. These constraints remain relatively high and closely consistent with the standard $\Lambda$CDM values inferred from CMB observations. As a result, the MEDE model does predict lower values of the $S_8$ parameter and therefore fails to resolve the tension with weak lensing measurements, which generally favor lower values of $S_8$~\cite{abbott2022dark,asgari2021kids}. It is worth noting that recent observational studies suggest that the $S_8$ tension is still under discussion and is not yet firmly established~\cite{abbott2023y3}. This indicates that, similarly to the $H_0$ tension, the MEDE model is unable to simultaneously resolve the current cosmological tensions within the observational framework considered here.

In Fig.~\ref{fig_3}, we present the redshift evolution of $H(z)/(1+z)$ (first row), $q(z)$ (second row), $w(z)$ (third row), and $f_{\mathrm{DE}}(z)$ (fourth row) for the MEDE and $\Lambda$CDM models. The solid curves shows the mean evolution, while the darker and lighter shaded regions represent the $68\%$ ($1\sigma$) and $95\%$ ($2\sigma$) confidence intervals, respectively. The first row shows the evolution of $H(z)/(1+z)$ for both the MEDE and $\Lambda$CDM models, which are compared with the cosmic chronometer (CC) measurements reported in Refs.~\cite{moresco2012improved,moresco2015raising,moresco20166}. For each dataset combination, it can be clearly seen in the zoomed-in panels of each plot that the MEDE model predicts slightly larger values than the $\Lambda$CDM model, although these deviations remain below the $0.5\sigma$ level.

In the second row, we present the evolution of the deceleration parameter $q(z)$ for the MEDE and $\Lambda$CDM models. It can be seen that both cosmological models show a smooth transition from an early decelerating epoch to the present accelerated expansion phase. For the DESI DR2 + CMB dataset combination, the MEDE model predicts slightly stronger late-time acceleration than the $\Lambda$CDM model. However, after adding the Pantheon$+$, DES-Dovekie, and Union3 datasets, the present-day values of the deceleration parameter ($q_0$) for both models become very close to each other.

In the third row, we show the evolution of the $w(z)$ parameter for the MEDE model, while $\Lambda$CDM assumes a constant value $w = -1$.  In the case of the CMB + DESI DR2 dataset combination, the mean evolution of $w(z)$ remains below $-1$. Indeed, the GEDE model exhibits a full phantom behavior ($w < -1$ at all epochs), consistent with $w_0 < -1$ and $w_0 + w_a < -1$~\cite{caldwell2002phantom}. However, after including the Pantheon$+$, DES-Dovekie, and Union3 datasets, the evolution of $w(z)$ shifts above the cosmological constant boundary $w=-1$, indicating a full quintessence behavior ($w > -1$ at all epochs), corresponding to $w_0 > -1$ and $w_0 + w_a > -1$~\cite{ratra1988cosmological}. The fourth column shows the evolution of $f_{\mathrm{DE}}(z)$ as a function of redshift. For all four dataset combinations, the dark energy function converges to $f_{\mathrm{DE}}(z)=1$ at the present epoch ($z=0$).

In Fig.~\ref{fig_2}, we show the 2D marginalized confidence contours at 68\% ($1\sigma$) and 95\% ($2\sigma$) confidence levels in the ${\alpha-\Omega_m}$ plane for the MEDE model. For the CMB + DESI DR2 dataset combination, the MEDE model yields $\alpha = 0.27^{+0.24}_{-0.21}$, which deviates from the $\Lambda$CDM reference value of $\alpha = 0$ at the $\sim 1.2\sigma$ level. When the Pantheon$+$ sample is combined with CMB and DESI DR2, the MEDE model yields $\alpha = -0.097 \pm 0.17$, corresponding to a deviation of about $0.57\sigma$ from the $\Lambda$CDM reference value. Similarly, for the CMB + DESI DR2 + DES-Dovekie combination, we obtain $\alpha = -0.14 \pm 0.16$, corresponding to a deviation of approximately $0.88\sigma$. For the CMB + DESI DR2 + Union3 dataset, we find $\alpha = -0.11^{+0.22}_{-0.19}$, corresponding to a deviation about $0.5\sigma$. These results indicate that, after including the SNe~Ia datasets, the parameter $\alpha$ remains fully consistent with the standard $\Lambda$CDM scenario, with deviations from $\alpha = 0$ remaining below the $1\sigma$ level and therefore showing no statistically significant evidence for departures from $\Lambda$CDM. Furthermore, the negative values of $\alpha$ correspond to the injection of energy at earlier redshifts.

In Fig.~\ref{fig_stat}, we show a statistical comparison between the MEDE and $\Lambda$CDM models for the different dataset combinations using $\Delta\chi^2_{\rm MAP}$, $\Delta{\rm AIC}$, and the logarithmic Bayes factor, $\ln\mathcal{B}_{\mathrm{MEDE},\,\Lambda\mathrm{CDM}}$. The upper panel shows the difference between the minimum $\chi^2$ values of the $\Lambda$CDM and MEDE models. We find that the CMB + DESI combination gives $\Delta\chi^2_{\rm MAP}\simeq0.002$, indicating that the two models provide an almost identical fit to the data. However, after including the Type Ia supernova datasets, the value of $\Delta\chi^2_{\rm MAP}$ increases to approximately $4$--$5$ units, showing that the MEDE model consistently achieves a lower minimum $\chi^2$ than the $\Lambda$CDM model for these dataset combinations and therefore provides a better fit to the observational data.

The middle panel shows the difference in the Akaike Information Criterion, $\Delta{\rm AIC}$. We find that the CMB + DESI combination gives $\Delta{\rm AIC}\simeq-1.99$, indicating a slight preference for the $\Lambda$CDM model. This is because both models provide an almost identical fit to the data ($\Delta\chi^2_{\rm MAP}\simeq0.002$), and therefore the additional free parameter of the MEDE model is not sufficiently justified by the improvement in the fit. Further adding the Type Ia supernova datasets gives positive values of $\Delta{\rm AIC}$, ranging from approximately $2.15$ to $3.29$. This shows that the better fit of the MEDE model is enough to compensate for its additional free parameter. As a result, the MEDE model is mildly preferred over the $\Lambda$CDM model for these dataset combinations.

The lower panel shows the logarithmic Bayes factor, $\ln\mathcal{B}_{\mathrm{MEDE},\,\Lambda\mathrm{CDM}}$, which quantifies the statistical preference between the MEDE and $\Lambda$CDM models. We obtain $\ln\mathcal{B}_{\mathrm{MEDE},\,\Lambda\mathrm{CDM}}=-0.17$ for the CMB + DESI DR2 dataset combination, $-0.02$ for CMB + DESI DR2 + Pantheon$+$, $-0.61$ for CMB + DESI DR2 + DES-Dovekie, and $-0.27$ for CMB + DESI DR2 + Union3. According to the revised Jeffreys' scale, all of these values correspond to inconclusive evidence in favour of the MEDE model over the $\Lambda$CDM model.

In this work, we investigated the MEDE model using the DESI DR2 dataset together with CMB information from the {\tt LoLLiPoP} (low-$\ell$) and {\tt HiLLiPoP} (high-$\ell$) likelihoods based on the latest Planck NPIPE PR4 data release, along with the Planck PR4 lensing and ACT DR6 lensing likelihoods, as well as three different Type Ia supernova datasets (Pantheon$+$, DES-Dovekie, and Union3). Our analysis shows that the MEDE model remains largely consistent with the $\Lambda$CDM model. Although the MEDE model shows small deviations from the $\Lambda$CDM model, the inclusion of the Pantheon$+$, DES-Dovekie, and Union3 datasets drives the model parameters toward values that are fully consistent with $\Lambda$CDM within the $1\sigma$ confidence level. In particular, the parameter $\alpha$ does not show any statistically significant deviation from $\alpha = 0$, suggesting no strong observational evidence for departures from the standard cosmological model.

We further find that the MEDE model does not provide a significant improvement to the $H_0$ tension. The inferred values of $H_0$ remain noticeably lower than the SH0ES measurement, while the sound horizon scale $r_d$ stays nearly unchanged compared to the $\Lambda$CDM prediction. Similarly, although the MEDE model predicts slightly lower values of the clustering parameter $S_8$, the reduction is not enough to ease the tension with weak lensing observations. Therefore, within the observational framework considered in this work, the MEDE model is unable to simultaneously resolve the current $H_0$ and $S_8$ cosmological tensions.

The dynamical evolution of the cosmological quantities also reveals that the MEDE model closely tracks the behavior of $\Lambda$CDM across most of cosmic history. Small deviations are observed in the evolution of $H(z)/(1+z)$ and the deceleration parameter $q(z)$, but these remain below the $0.5\sigma$ level. Moreover, the evolution of the equation of state parameter $w(z)$ indicates that the CMB + DESI DR2 dataset combination favors a full phantom-like behavior, whereas the inclusion of SNe~Ia datasets shifts the evolution toward a full quintessence-like regime. Finally, the statistical analysis shows that the MEDE model provides a better fit to the observational data than the $\Lambda$CDM model, as indicated by the positive values of $\Delta\chi^2_{\rm MAP}$ and, for the combinations including Type Ia supernova data, positive values of $\Delta{\rm AIC}$. Nevertheless, the logarithmic Bayes factor indicates only inconclusive evidence in favour of the MEDE model over the $\Lambda$CDM model according to the revised Jeffreys' scale. Despite this overall consistency, the MEDE model does not show the phantom crossing suggested by the DESI DR2 results~\cite{gu2025dynamical}. Therefore, our analysis indicates that, although the MEDE model remains compatible with $\Lambda$CDM, it does not support the specific dynamical dark energy behavior suggested at by DESI DR2 observations.

\bibliographystyle{elsarticle-num}
\bibliography{mybib.bib}

\end{document}